\documentclass{article}
\usepackage[utf8]{inputenc}
\usepackage[a4paper,
            bindingoffset=0.2in,
            left=1.2in,
            right=1.2in,
            top=1.2in,
            bottom=1.2in,
            footskip=.25in]{geometry}
\usepackage{titling}
\usepackage[colorlinks = true,
            linkcolor = blue,
            urlcolor  = blue,
            citecolor = blue,
            anchorcolor = blue]{hyperref}

\newcommand{\Mod}[1]{\mathrm{mod}\ #1}

\usepackage{setspace} 
\usepackage{etoolbox}
\AtBeginEnvironment{quote}{\par\singlespacing}

\usepackage{authblk} 

\usepackage{tikz} 
\usepackage{caption}
\usetikzlibrary{quantikz}
\usepackage{pgfplots,pgfplotstable}

\DeclareFixedFont{\ttb}{T1}{txtt}{bx}{n}{12} 
\DeclareFixedFont{\ttm}{T1}{txtt}{m}{n}{12}  

\usepackage{color}
\definecolor{deepblue}{rgb}{0,0,0.5}
\definecolor{deepred}{rgb}{0.6,0,0}
\definecolor{deepgreen}{rgb}{0,0.5,0}

\usepackage{listings}

\usepackage{mathtools}

\newcommand\pythonstyle{\lstset{
language=Python,
basicstyle=\ttm,
morekeywords={self},              
keywordstyle=\ttb\color{deepblue},
emph={MyClass,__init__},          
emphstyle=\ttb\color{deepred},    
stringstyle=\color{deepgreen},
frame=tb,                         
showstringspaces=false
}}

\lstnewenvironment{python}[1][]
{
\pythonstyle
\lstset{#1}
}
{}

\newcommand\pythoninline[1]{{\pythonstyle\lstinline!#1!}}

\title{
Making Music Using Two Quantum Algorithms}
\author[1,2,*]{Euan J. Allen}
\author[1]{Jacob F. F. Bulmer}
\author[3]{Simon D. Small}
\affil[1]{\small Quantum Engineering Technology Labs, H. H. Wills Physics Laboratory and Department of Electrical \& Electronic Engineering, University of Bristol, BS8 1FD, United Kingdom}
\affil[2]{Centre for Photonics and Photonic Materials, Department of Physics, University of Bath, Bath, BA2 7AY, United Kingdom}
\affil[3]{Tunnel of Reverb, \url{https://www.tunnelofreverb.com/}}
\affil[*]{ea901@bath.ac.uk}
\date{\today}

\begin{document}
\begin{titlingpage}
    \centering
    
    \maketitle
    \begin{abstract}
This document explores how to make music using quantum computing algorithms. The text is an unedited pre-publication chapter which will appear in the book “Quantum Computer Music”, Miranda, E. R. (Editor). This chapters provides the background and specific details of a collaboration formed in 2021 between the Quantum Engineering Technology Labs - a quantum computing and technology research group at the University of Bristol - and music artist, producer and audio engineer Simon Small. The goal of the collaboration was to explore how the data and concepts used in the research at the university could be `sonified' to create sounds or even make music. \\

The audio outcomes of the collaboration can be heard at the following reference~\cite{QLquantummusic} or at \url{https://bohmelectronic.bandcamp.com/}. Reference~\cite{QLquantummusic} additionally includes the lab audio samples and sheet music for other musicians to utilise the data and sounds generated from this work. \\ \vspace{0.5cm}
        \centering
        \begin{figure}[h]
            \centering
            \includegraphics[width=0.55\textwidth]{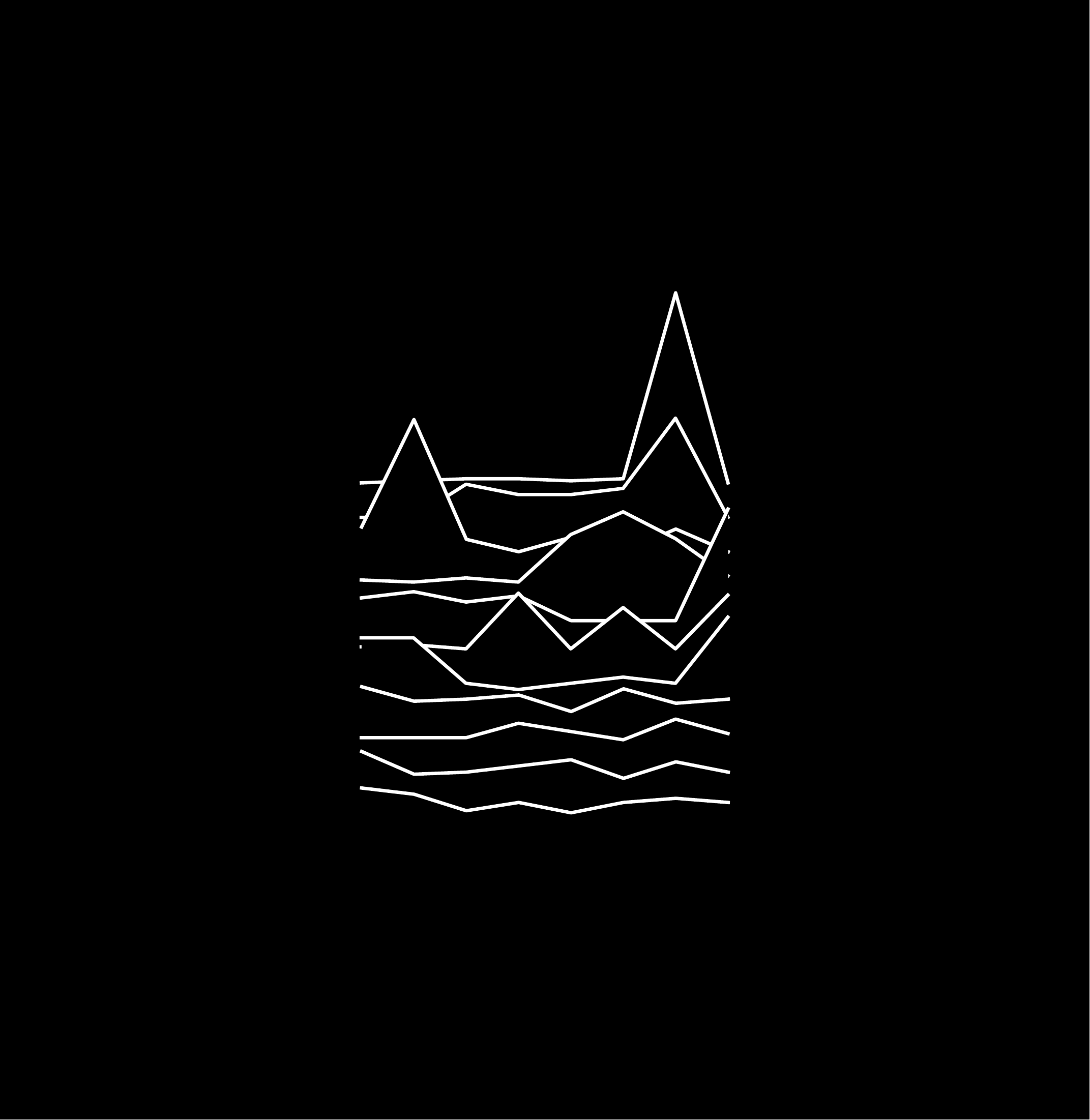}
            \caption{A pictorial representation of Grover's algorithm on a 3-qubit processor, plotted in the style of the Joy Division album cover ``Unknown Pleasures''.}
        \end{figure}

    \end{abstract}
\end{titlingpage}




\section{Introduction}
Computers have and continue to shape the sound and music landscape that we experience. Whether that be as tools to make music with or inspiration for writing and composing music, the impact of computation in sound generation is not difficult to find. 

Quantum computers are a new type of device that completes computational tasks in a very different way to the (classical) computers we experience in everyday life. Utilising aspects of quantum physics, such as entanglement and superposition, quantum computers are able to solve problems which are very difficult for a classical computer to complete. 

The manner in which quantum computers work is distinct from classical computers. It is therefore of interest to explore what impact this technology could have on future sound and music generation, much like the impact conventional computers had in the early 20th century.  Conversely, it is also of interest to explore how sound and music might help in disseminating concepts of quantum physics and computing to a wider audience - an audience that will likely feel the impact of this technology at some point in the future. 

Quantum computers work and process quantum information, typically stored in quantum bits (qubits), whilst classical machines process classical information, typically stored in binary bits taking a value of 0 or 1. Whilst there are examples of quantum versions of sound or acoustic waves (phonons)~\cite{satzinger2018quantum}, all music and sound that is processed by the human ear in day-to-day life can be understood as being entirely classical. It is an interesting task to work out which quantum concepts or computational algorithms have a sufficiently interesting classical output at the end of the computation to make them audially distinct or interesting from a classical machine. This is part of the task of exploring quantum computing music.

This chapters provides the background and specific details of a collaboration formed in 2021 between the Quantum Engineering Technology Labs~\cite{qetlabs} - a quantum computing and technology research group at the University of Bristol - and music artist, producer and audio engineer Simon Small~\cite{tunnelofreverb}. The goal of the collaboration was to explore how the data and concepts used in the research at the university could be `sonified' to create sounds or even make music. 

The project focused on two key concepts for sound generation: quantum random walks and Grover's search algorithm. These two quantum algorithms formed the basis for sound and music generation, the culmination of which resulted in two full musical compositions. This chapter is split in to three key sections. The first two provide a technical introduction to both random quantum walks and Grover's algorithm, covering how the algorithms work and how to produce data from them that can be used to generate sound. The final section covers how to take this data and use it for musical composition. Details of other techniques used in the collaboration, such as audio samples from the laboratory, are also detailed in the final section. 

The audio outcomes of the collaboration can be heard at the following references~\cite{QLquantummusic,QLbandcamp}. These links include the final pieces, audio samples, and a full musical pack for other musicians to utilise the data and sounds generated from this work.

\section{Random Melodies from Quantum Walks}

A random walk is a great example of a place where there is a clear difference in the behaviour of quantum and classical physics~\cite{aharonov1993quantum}.
It is therefore frequently used as a thought experiment to teach the concepts of quantum mechanics, but has also been used to inspired quantum algorithms~\cite{childs2003exponential}, and has been realised in experiments~\cite{schreiber2010photons}.
This section investigates how we can use the difference between classical and quantum random walks to create musically contrasting melodies.

We first describe how we define a quantum random walk, and show how measurement can lead to classical dynamics.
We then discuss how we simulate these systems.
Although there are a wide variety of excellent quantum simulation libraries, we chose to implement our simulation using only Python and NumPy.
This allows us to clearly see how all of our operations are defined and implemented.

\subsection{Quantum Random Walks}

We will define our random walk as follows, imagine that every time you want to take a step, you toss a coin.
If it lands on heads, you step to the left.
If it lands on tails, you step to the right.
We add a further rule of a \textit{periodic boundary condition}, which can be described by saying that we have a fixed number of sites to stand on, if you reach the left edge, and need to step to the left, you move to the right edge and vice versa. This is equivalent to arranging our sites in a circle where if you go all the way around the circle you end up back where you started. 

We wish for our sites to neatly correspond to musical notes in a scale, so we choose to use 14 sites, labelled from 0 to 13, which can then be turned into notes from 2 octaves of an 8-note scale.

If we assume that the state of our coin, and the site our walker is standing on are prepared independently from each other, our quantum system can be defined like:

\begin{equation}
    \ket{\psi} = \ket{C} \otimes \ket{S}
\end{equation}

where $\ket{C}$ is the wavefunction describing the coin and $\ket{S}$ is the wavefunction describing the site.
The symbol ``$\otimes$'' is a tensor product, which here we can just think of as a symbol which shows that we are taking the two systems $\ket{C}$ and $\ket{S}$ and thinking of them as one system.
We define coin basis states: $\ket{\text{heads}}$ and $\ket{\text{tails}}$, and site basis states $\ket{j}$ where $j$ can take any value from 0 to 13 ($j \in \{0,1,2,\dots, 13\}$).

The operation we apply to simulate a coin toss is a Hadamard rotation. 
This maps the coin states like: 
\begin{align}
    \ket{\text{head}} & \to \frac{\ket{\text{heads}} + \ket{\text{tails}}}{\sqrt{2}} \\ 
    \ket{\text{tails}} & \to \frac{\ket{\text{heads}} - \ket{\text{tails}}}{\sqrt{2}}.
\end{align}

These two states are equal superposition states between heads and tails, with a phase difference between the two. We also need an operation which will move the site of the walker, depending on the outcome of the coin. We call this the \textit{move} operator, $\hat{M}$:

\begin{align}
    \hat{M} = & \ket{\text{heads}}\! \bra{\text{heads}} \otimes \sum_{j=0}^{13} \ket{(j + 1) \Mod{14}}\! \bra{j} \\
    & + \ket{\text{tails}}\! \bra{\text{tails}} \otimes \sum_{j=0}^{13} \ket{(j-1) \Mod{14}}\!\bra{j}.
\end{align}

By inspecting this equation, we can see that when this operator is applied to a state $\ket{\psi}$, the walker will move to an adjacent site.
For a term with containing a $\ket{\text{heads}}$ coin state, this will be a site to the right ($+1$) and for $\ket{\text{tails}}$, it the left ($-1$). 

Also notice the ``$\Mod{14}$'' operation.
This modulo division means that if we try to step to position $j=14$, we move to $j = 0$, since $14\ \Mod{14} = 0$.
Likewise, if try to move to $j=-1$, we move to $j=13$.
This is what allows us to encode the periodic boundary and put the sites in the circle configuration.

To implement our quantum walk, we alternate between applying Hadamard operations to the coin, and then the move operation to our coin and walker system. The move operation generates entanglement between the site of the walker and the state of the coin. 
Eventually, the state $\ket{\psi}$ will have support across the whole Hilbert space, which exists in $2 \times 14 = 28$ dimensions.

As an example, we will work out the state of the system explicitly after the first steps. 
Let us imagine that we started in position $j=7$, and we initialise our coin in $\ket{\text{heads}}$.
\begin{equation}
    \ket{\psi} = \ket{\text{heads}} \otimes \ket{7}
\end{equation}
We apply the Hadamard operation to the coin:
\begin{equation}
    \hat{H} \otimes \mathbf{1} \ket{\psi} = \frac{(\ket{\text{heads}} +\ket{\text{tails}}) \otimes \ket{7}}{\sqrt{2}} 
\end{equation}
where $\mathbf{1}$ is the identity operation (i.e. the do-nothing operation) acting on the walker.
Then we apply the move operation to the system:
\begin{equation}
    \hat{M} (\hat{H} \otimes \mathbf{1}) \ket{\psi} = \frac{\ket{\text{heads}}\otimes\ket{8} +\ket{\text{tails}} \otimes \ket{6}}{\sqrt{2}} .
    \label{one_round}
\end{equation}

This state is now \textit{entangled}, as we can no longer consider the coin and the walker state separately. 
If we were to measure position of the walker, and see it was at position 8, we would know that the coin was showing heads.

If we apply the coin flip then the move operation several times before we measure, we would have a much more complicated state.
However, all of the terms in the state would contain either a heads of a tails outcome. 
If we want to know the probability of the walker being in a particular position, but we do not care about the state of the coin, we can sum together the heads and tails probabilities associated with that position:
\begin{equation}
    p(j) = p(j, \text{heads}) + p(j, \text{tails}),
\end{equation}
where $p(j)$ is the probability of the walker being in position $j$, and $p(j, \text{heads/tails})$ is the probability of a walker being in position $j$ with the coin in a state heads/tails respectively.

To get these probabilities, we just have to apply the \textit{Born rule}. 
This tells us that the probabilities of measuring a quantum state is given by the absolute-square of its amplitude in the state.

For example, for the state in equation~\ref{one_round}:
\begin{align}
    p(j=6) & = p(j=6, \text{heads}) + p(j=6, \text{tails}) \\
    & = 0 + \left| \frac{1}{\sqrt{2}} \right|^2 \\
    & = 1/2
\end{align}

However, an important fact of quantum mechanics is that we disturb a system when we measure it. 
If we perform a measurement of the position of the walker, we collapse it to the result of the measurement.
In the case above, we would end up with $\ket{S} = \ket{6}$.

So, we had a $50\%$ chance of seeing the tails outcome and moving to the left. 
We see this movement, and we now know that we are in position 6. 
If we consider repeating this process every step, we see that measurement at each step collapses our random walk into completely classical dynamics.

We wish to also probe the quantum dynamics of our system, and so we give ourselves some abilities which cannot be achieved in real-world quantum experiments: we let ourselves look \textit{inside} the state.
In simulation, we can see the probability of a given measurement from occurring without needing to perform the measurement.
It is this ability which we harness to generate data for our quantum walk.

We could actually replicate the dynamics of looking inside the state in an experiment by making some changes to the rules. 
Here, what we would have to do is after each measurement, we start the walk again from the beginning.
Before we make the next measurement, we let the walker take one more step than in for the previous measurement before we take the next measurement.

\subsection{Simulation}
In this section, we introduce how to construct the operations we defined in the previous section using Python and NumPy, both of which are free to use software packages.
 
As a first step, we import Numpy, define the number of sites, the starting position and the Hadamard operation, as a $2 \times 2$ unitary matrix:
 
\begin{python}
import numpy as np
sites = 14
starting_position = 7
H = np.array([[1,1],[1,-1]]) / np.sqrt(2)

# tensor product with identity
H_I = np.kron(H, np.identity(sites))
\end{python}

We can define a unitary which moves the coin to the right and left:
\begin{python}
U_right = np.diag(np.ones(sites-1), -1)
U_right[0, sites-1] = 1

U_left = np.diag(np.ones(sites-1), 1)
U_left[sites-1,0] = 1
\end{python}
To define the move operator, we also need to define a matrix which selects the heads or tails outcomes:
\begin{python}
heads = np.array([1,0])
tails = np.array([0,1])

heads_heads = np.outer(heads, heads)
tails_tails = np.outer(tails, tails)

move = (np.kron(heads_heads, U_right) 
    + np.kron(tails_tails, U_left))
\end{python}
We can also initialise our state:
\begin{python}
C = heads
S = np.zeros(sites)
S[starting_position] = 1
psi = np.kron(C, S)
\end{python}
We are now ready to simulate our random walk.
\begin{python}
classical = False
steps = 1000
for i in range(steps):
    psi = H_I @ psi
    psi = move @ psi
    # measure position of walker
    probs = abs(psi[:sites])**2 + abs(psi[sites:])**2
    pos = np.random.choice(range(sites), p=probs)
    print(pos)
    # if classical, we must project our state 
    # onto the measured outcome
    if classical:
        S = np.zeros(sites)
        S[pos] = 1
        C = heads
        psi = np.kron(C, S)
\end{python}
If we set the variable \verb|classical| to \verb|True|, then we see that we collapse our system after each measurement to the observed outcome, recovering the classical dynamics.

When we view the strings of numbers which this program returns, we see that both classical and quantum start returning numbers close to the starting position, gradually drifting apart. 
The classical version only can take one step to the left or right each time, whereas the quantum version can take larger steps. 
This is reminiscent of how a quantum wavefunction spreads out over time.
Importantly, this creates clearly noticeable differences to the melodies created from the data.

\section{Grover's Algorithm}
Grover's search algorithm is one of the most well know algorithms that can be run on a quantum computer. It was first introduced by Lov Kumar Grover in 1996/7~\cite{grover1996fast, grover1997quantum} and tackles a problem known as an `unstructured search'. We introduce this specific problem in Section~\ref{sec:unstructuredsearch} and introduce how a quantum computer may offer an improvement in implementing such an algorithm. The specific details of the algorithm and how it can be implemented on a quantum computer are discussed in Section~\ref{sec:groverstructure}. Data outputs are provided from the algorithm at various points.

\subsection{An Unstructured Search}
\label{sec:unstructuredsearch}
The problem that Grover's algorithm can be applied to is that of searching through a list or database of entries to find a particular value of interest. An example would be searching for a particular person in a phone book. In particular, the problem of interest is applied to an \textit{unstructured} list, whereby the entries in the phone book are listed randomly (rather than, for example, in alphabetical order). Whilst this type of problem seems a simple one, searching through database is a fundamental feature of many computational tasks and most of that data is unsorted~\cite{gandomi2015beyond}. 

In order to be able to say that the quantum (Grover's) algorithm is better than the classical one, we need to assess how difficult it is to complete the unstructured search task classically. If we consider a phone book with $N$ entries in a completely random order, then the task is to search through the book to find a particular name. A simple method would be to just go through the list entries, searching until you find the name you want. If $N$ is large, then the initial probability of finding the entry you want is very small. As you check more and more entries, the probability of finding the one you want increases. Eventually, after checking half the entries ($N/2$) you will have a 50\% chance of finding the name you want. That's to say, that if you want a 50\% chance of finding the correct entry, you will have to check $N/2$ times. 

Note that if we double the size of our phone book or database, then the number of entries we have to check to get a 50\% probability also doubles. Similarly, if the phone book gets a million times larger, it takes a million times longer to find the result. This is an indication of how the algorithm \textit{scales} and is predicted entirely by the dependence of the algorithm on $N$ (in this case, a linear dependence). Sometimes, the scaling of the algorithm with $N$ is the only bit of information that is relevant to the comparison. In this case, the algorithm is said to scale as $\mathcal{O}(N)$ or `with a leading order linear in $N$'. Note that whilst we have discussed here a very particular method of searching the database classically, one actually finds that any classical algorithm will complete the search task $\mathcal{O}(N)$~\cite{grover1996fast}. 

In contrast to classical methods, Grover derived an algorithm using a quantum computer that is able to complete the task with only $\mathcal{O}(\sqrt{N})$ checks in the phone book or accesses to the database~\cite{grover1996fast, grover1997quantum}. What this means is that if the phone book doubles in size, then the quantum algorithm only takes $\sqrt{2}$ longer, or about 1.4 times. If it goes up by a factor of one million, then the quantum algorithm only needs one thousand times longer. If we assume that to check each name in the phone book takes a certain fixed amount of time, then the quantum algorithm will be able to search much faster than the classical one - particularly for very large phone books. Even in the case where the quantum algorithm is slower at checking the phone book entries than the classical one, there will always be a value of $N$ (a size of phone book) where the quantum algorithm outperforms the classical one and can complete the task is less time. 

\subsection{Structure of the algorithm}\label{sec:groverstructure}
This section will give a surface level overview of Grover's algorithm and how it is implemented on a quantum computer. We will give particular focus to aspects of the algorithm that are relevant to later sections. A more detailed description of how the algorithm works can be found in references~\cite{grover1996fast, grover1997quantum,nielsen2002quantum}.

Suppose we want to search through a list of eight entries, which we label using binary encoding: 000, 001, 010, 011, 100, 101, 110, 111 - which encode values from 0 to 7. These can be encoded using three qubits $\ket{q_1 q_2 q_3}$ with $q_1$, $q_2$, or $q_3$ taking a value of $0$ or $1$. The goal of the quantum algorithm is to take in a set input qubit register of qubits in the state $\ket{000}$, perform some kind of manipulation and computation on the qubits, and then (ideally) output the qubit values which correspond to the entry we are searching for. For this example we consider the entry we are searching for to be $\ket{110}$, or the seventh entry, but in principle the algorithm works for any of the eight possible outcomes.

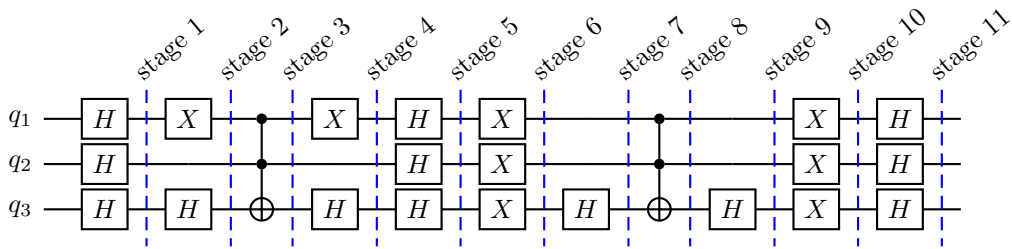
\begin{figure}[h]
\centering
\begin{quantikz}[row sep={0.6cm,between origins},slice all,slice
titles=stage \col,slice style=blue,slice label style
={inner sep=1pt,anchor=south west,rotate=40}]
\lstick{$q_1$} & \gate{H} & \gate{X} & \ctrl{1} & \gate{X} & \gate{H} & \gate{X} & \qw & \ctrl{1} & \qw & \gate{X} & \gate{H} & \qw \\
\lstick{$q_2$} & \gate{H} & \qw & \ctrl{1} & \qw & \gate{H} & \gate{X} & \qw & \ctrl{1} & \qw & \gate{X} & \gate{H} & \qw \\
\lstick{$q_3$} & \gate{H} & \gate{H} & \targ{} & \gate{H} & \gate{H} & \gate{X} & \gate{H} & \targ{} & \gate{H} & \gate{X} & \gate{H} & \qw
\end{quantikz}
\caption{A three qubit Grover's algorithm implementation. Each horizontal line (or 'rail') corresponds to an individual qubit with boxes on that rail representing quantum gates acting on each qubit. The gates displayed include the Hadamard (H), Pauli X (X), and control-NOT (control node $\bullet$, 'NOT' node $\bigoplus$). The algorithm runs from left to right.\label{fig:groverGates}}
\end{figure}

The operations performed on the qubits during any algorithm are defined by quantum gates. The arrangement of quantum gates for Grover's algorithm are shown in Figure~\ref{fig:groverGates}. Each horizontal line or `rail' corresponds to a qubit ($q_1$, $q_2$, and $q_3$), with operations or gates defined by any square or circle drawn on the rail, running in order from left to right. In this case, the quantum gates required to perform the algorithm include the Hadamard ($H$), Pauli X ($X$), and control-NOT (control node $\bullet$, `NOT' node $\bigoplus$) gate. These gates in total perform the algorithm as desired. A more detailed description of why this arrangement of quantum gates performs Grover's algorithm can be found in the following references~\cite{grover1996fast, grover1997quantum,nielsen2002quantum,qiskitGrover}.

As we have stated previously, at the start of the algorithm the qubits are initiated in the state $\ket{000}$. The first action of the algorithm is to apply a Hadamard gate ($H$) to each qubit. The action of this gate on an initial state $\ket{0}$ can be described as $\ket{0} \xrightarrow{H} \frac{\ket{0} + \ket{1}}{\sqrt{2}}$, and so a Hadamard generates an equal superposition of $\ket{0}$ and $\ket{1}$. When a Hadamard is applied to all three initial qubit states, the quantum state generated is an equal superposition across all possible qubit outcomes: $\ket{000} \xrightarrow{H} \frac{\ket{000} + \ket{001}+\ket{010}+...+\ket{111}}{\sqrt{8}}$. The amplitude of the quantum wavefunction is equally split between all possible measurement outcomes. Practically, what this means is that at stage 1 in the algorithm, any measurement of the qubits has an equal chance of outputting any of the eight possible binary output values. In a sense, at this point in the algorithm the particular value we are looking for is completely unknown. Asking the algorithm which entry in the list is the one we are looking for (by measuring the state of the qubits at this point), will randomly output any of the possible entries. 

The complete randomness of stage 1 is in complete contrast with the end of the algorithm (stage 11) which, as we have already stated, will ideally always output $110$ and therefore has all of it's quantum amplitude in the state $\ket{110}$. In practise, the end of the algorithm manipulates \textit{most} of the quantum amplitude to the state $\ket{110}$ because of small details in how the algorithm works or noise in the quantum machine completing the computation. It is the goal of the applied gates/the algorithm to redistribute the wavefunction to place the maximal amount in the correct output. This means that at the end of the computation, we have the greatest chance to measure the correct qubit output values and therefore find our entry. This contrast between the complete randomness of the start of the algorithm, and the complete (or near complete) determinism of the end of the algorithm is why it was of interest for music generation. The journey between these two stages is also of interest and gives insight to the inner workings of Grover's algorithm.


\subsection{Simulating Grover's Algorithm}
There are a number of publicly available tools to simulate Grover's algorithm. Here we focus on QuTech’s Quantum Inspire Home Platform~\cite{qutechQI} which allows users to simulate the algorithm and measure qubit outcomes after processing. Simulating the algorithm allows a user to generate lists of measurement outcomes at each stage in the algorithm which can then but used as source material for sound and music generation. 

\begin{figure}[h]
    \centering
\begin{tikzpicture}
\begin{axis}[
/pgf/number format/.cd,fixed,precision=5,
ybar,
enlargelimits=0.2,
legend style={at={(0.5,-0.15)},
anchor=north,legend columns=-1},
ylabel={Output Proportion},
height=5cm,
width=13cm,
bar width=25,
symbolic x coords={000,001,010,%
011,100,101,110,111},
xlabel={Qubit Output},
xtick=data,
nodes near coords,
nodes near coords align={vertical},
]
\addplot coordinates {
(000,0.18) (001,0.11) (010,0.12)
(011,0.14) (100,0.06) (101,0.17) (110,0.10) (111,0.12)
};
\end{axis}
\end{tikzpicture}
    \caption{Histogram plot of the qubit measurement outcomes for stage 4 in the algorithm as defined by Figure~\ref{fig:groverGates}. The numbers relate to the proportion of measurement outcomes of that particular value recieve for 100 measurements of the qubit state at stage 4. }
    \label{fig:outcomesStage3}
\end{figure}
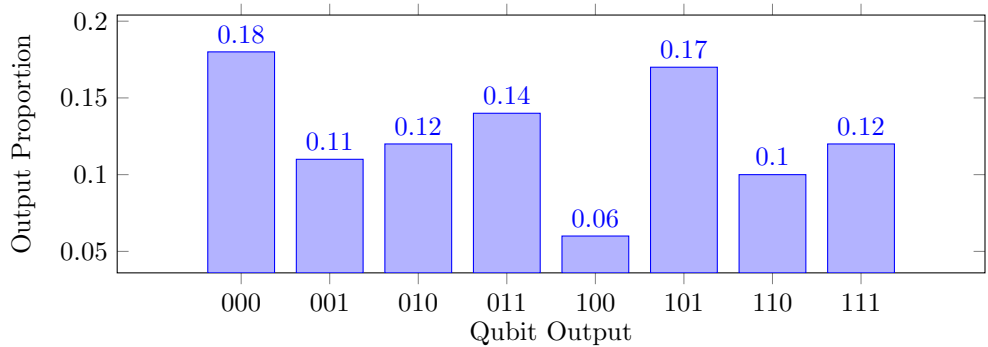

Each measurement of the qubit values is probabilistic, with the probability of measuring any particular outcome described by the wavefunction amplitude of that qubit outcome at that point in the algorithm. This means that preparing the qubits in the state $\ket{000}$, processing to a particular stage of the algorithm, and then measuring the qubits will produce a different output if you repeat it multiple times. For example, at stage 1, a set of ten measurements could produce the following outcome (where we have no converted back from binary to numerical values, e.g. 010 = 2):  1, 0, 3, 2, 4, 7, 2, 2, 6, 2. Similarly, a set of ten measurements at stage 11 of the algorithm produces the following output: 6, 6, 3, 5, 6, 6, 6, 6, 6, 6. These two lists demonstrate the contrast discussed before between the complete randomness of stage 1 and the (near) deterministic output of stage 11 - always outputting our correct entry value `6'.  By completing a tally of the proportion of each output we measure, each stage can also be displayed graphically. This is complete for 100 measurements of stage 4 in the algorithm in Figure~\ref{fig:outcomesStage3}.

\begin{figure}
\centering
\begin{tikzpicture}
\pgfplotstableread{
plot1   plot2   plot3   plot4   plot5   plot6   plot7   plot8   plot9   plot10  plot11
0       0       0       0       0       0       0       0       0       0       0
0.190   0.200   0.100   0.180   0.250   0.040   0.110   0.030   0.120   0.000   0.000   
0.160   0.090   0.100   0.110   0.250   0.050   0.140   0.020   0.640   0.000   0.010   
0.080   0.100   0.100   0.120   0.030   0.030   0.090   0.040   0.060   0.160   0.020
0.120   0.130   0.170   0.140   0.000   0.300   0.120   0.020   0.000   0.110   0.020
0.070   0.160   0.130   0.060   0.030   0.030   0.000   0.250   0.070   0.110   0.010
0.120   0.070   0.090   0.170   0.060   0.230   0.000   0.360   0.000   0.140   0.020
0.140   0.150   0.190   0.100   0.030   0.030   0.000   0.230   0.110   0.480   0.920
0.120   0.100   0.120   0.120   0.350   0.290   0.540   0.050   0.000   0.000   0.000
0       0       0       0       0       0       0       0       0       0       0
}\dummydata

\begin{axis}[
    xlabel=Qubit outcome,
    ylabel=Stage,
    zlabel=Measurement tally,
    samples=30,
    domain=-4:4,
    samples y=0, ytick={1,3,5,7,9,11},
    zmin=0,
    area plot/.style={
        fill opacity=0.75,
        draw=orange!80!black,thick,
        fill=orange,
        mark=none,
    }
]
\pgfplotsinvokeforeach{11,10,...,1}{
    \addplot3 [area plot] table [x expr=\coordindex-1, y expr=#1, z=plot#1] {\dummydata};
}
\end{axis}
\end{tikzpicture}
\caption{Two plots displaying the proportion of each qubit outcome measured at different stages of Grover's algorithm.  The data presented here is also displayed in Table~\ref{tab:grovermntoutcomes}.}
\label{fig:qubitoutcome}
\end{figure}
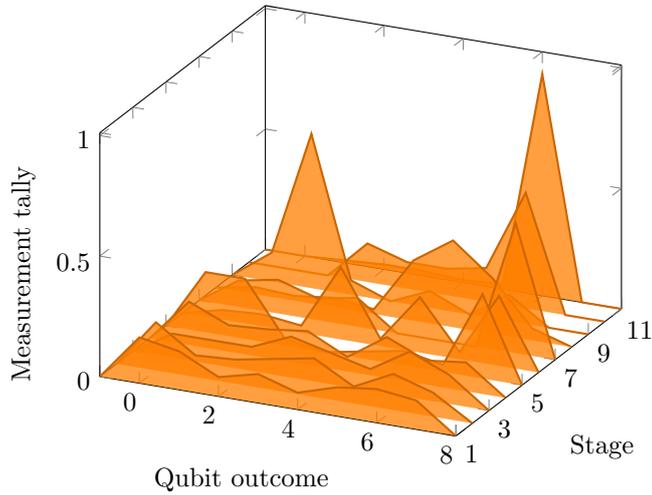

The Quantum Inspire Home Platform~\cite{qutechQI} allows one to simulate all stages of the algorithm and make multiple measurements of the state at each stage. Completing this process and making 100 measurements at each stage produces 11 lists of 100 numbers. The proportion of each measurement outcome at each stage of the algorithm completed using this platform is presented in Table~\ref{tab:grovermntoutcomes} and Figure~\ref{fig:qubitoutcome}. This list of numbers, documenting the state of the qubits at each point in the algorithm, can be used as a seed for sound and music generation. This is described in the following section.

\begin{table}[]
\centering
\begin{tabular}{c|cccccccc}
\multicolumn{1}{l|}{} & \multicolumn{8}{c}{\textbf{Proportion of Measurement Outcomes}}                                                                 \\
\textbf{Stage}        & \textbf{000} & \textbf{001} & \textbf{010} & \textbf{011} & \textbf{100} & \textbf{101} & \textbf{110} & \textbf{111} \\ \cline{2-9} 
\textbf{1}            & 0.19         & 0.16         & 0.08         & 0.12         & 0.07         & 0.12         & 0.14         & 0.12         \\
\textbf{2}            & 0.20         & 0.09         & 0.10         & 0.13         & 0.16         & 0.07         & 0.15         & 0.10         \\
\textbf{3}            & 0.10         & 0.10         & 0.10         & 0.17         & 0.13         & 0.09         & 0.19         & 0.12         \\
\textbf{4}            & 0.18         & 0.11         & 0.12         & 0.14         & 0.06         & 0.17         & 0.10         & 0.12         \\
\textbf{5}            & 0.25         & 0.25         & 0.03         & 0.00         & 0.03         & 0.06         & 0.03         & 0.35         \\
\textbf{6}            & 0.04         & 0.05         & 0.03         & 0.30         & 0.03         & 0.23         & 0.03         & 0.29         \\
\textbf{7}            & 0.11         & 0.14         & 0.09         & 0.12         & 0.00         & 0.00         & 0.00         & 0.54         \\
\textbf{8}            & 0.03         & 0.02         & 0.04         & 0.02         & 0.25         & 0.36         & 0.23         & 0.05         \\
\textbf{9}            & 0.12         & 0.64         & 0.06         & 0.00         & 0.07         & 0.00         & 0.11         & 0.00         \\
\textbf{10}           & 0.00         & 0.00         & 0.16         & 0.11         & 0.11         & 0.14         & 0.48         & 0.00         \\
\textbf{11}           & 0.00         & 0.01         & 0.02         & 0.02         & 0.01         & 0.02         & 0.92         & 0.00        
\end{tabular}
\caption{\normalfont{The proportion of measurement outcomes generated for 100 measurements of the qubit state at all 11 stages of Grover's algorithm.}}
\label{tab:grovermntoutcomes}
\end{table}

\label{sec:groversmusic}
\newpage
\section{Making Music Using Quantum Algorithms}

In this section, the process of taking raw data from Grover’s algorithm and the random quantum walk and translating it into an appropriate format for making music is discussed. Section~\ref{sec:music1} explains how the MAX MSP programming language turns this data into MIDI and then transmits it for musical hardware and software to interpret. Section~\ref{sec:music2} discusses the musical nature of the data and the audible differences between the algorithms. Section~\ref{sec:music3} features a quote from the composer, Simon Small, about the composition process and he discusses the emotional reaction to the quantum-data driven melodies.

\subsection{Raw Data and Processing into MIDI}\label{sec:music1}

\begin{figure}[h]
    \centering
    \includegraphics[width=\textwidth]{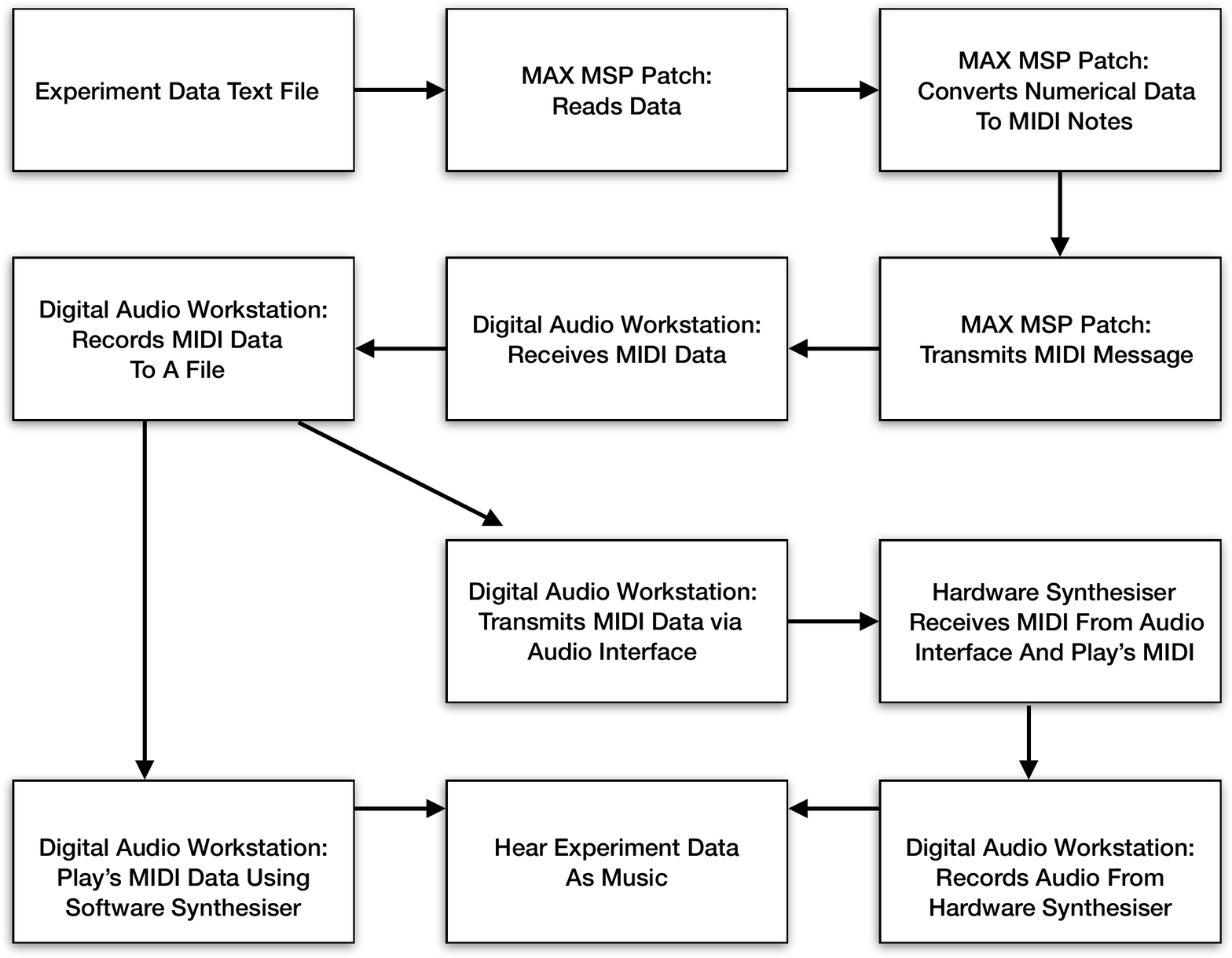}
    \caption{Workflow diagram of the processing of quantum data into a format that can be used for sound generation.}
    \label{fig:flowchat}
\end{figure}

Once raw data is simulated from the Random Walk and Grover algorithms, processing it into data that could be understood by musical hardware and software can begin. An obvious choice for this is the MIDI protocol, the universal standard for communicating data to hardware and software alike. This can be achieved using a patch in Cycling 74's MAX MSP, a visual programming language for music and multimedia. This process is discussed in the following text and summarised in Figure~\ref{fig:flowchat}.

\begin{figure}[h]
    \centering
    \includegraphics[width=\textwidth]{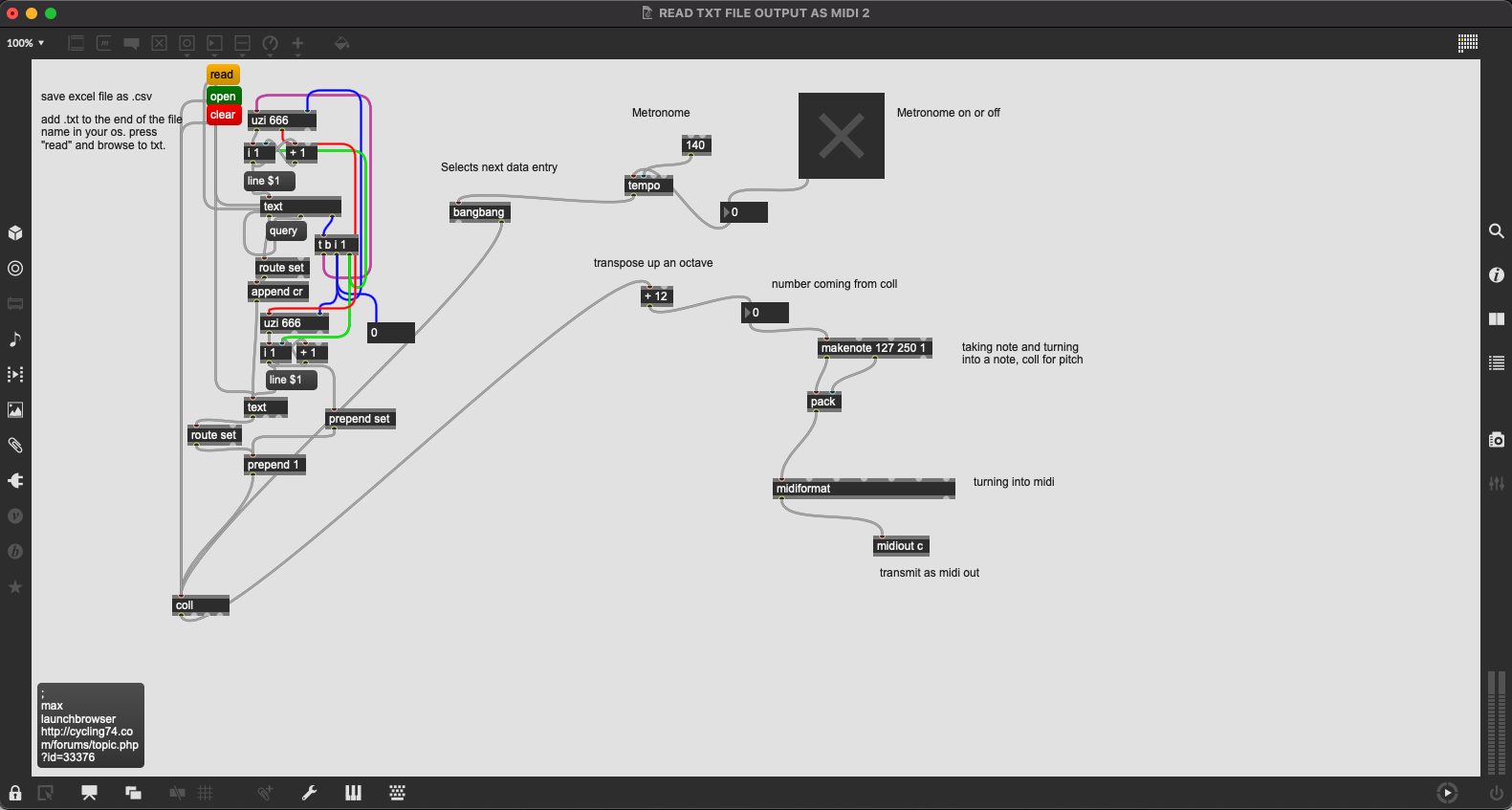}
    \caption{An example of the software inteface for the MAX MSP software which is used to take the raw data input and provide a MIDI output. }
    \label{fig:maxpatch}
\end{figure}

The patch works by importing a text file containing data from each algorithms simulation, cycling through the numbers at a set tempo and outputting them as MIDI data. A simple tempo object is used to drive the data from the text file, pushing it through the sequence of numbers at a set tempo. This can be started and stopped by the user using the X button (see Figure~\ref{fig:maxpatch}). The makenote object outputs a MIDI note-on message paired with a velocity value followed by a note-off message after a specified amount of time. In this patch, a fixed amount of time between note on and note off is used, but this can also be tempo relative. Using a fixed time allows the user to hear the experiments output as a fixed flow of notes. The numbers from the text files are input into this object to designate the pitch. It is worth noting that the pitched notes are transposed an octave up before hitting the makenote object as in their raw form they are extremely low and almost inaudible. A midiformat object was used to pack the data into a MIDI message, which is then transmitted to a specific MIDI port with midiout.

With the MIDI data transmitting from MAX MSP, the user can then route it as needed. Routing into the digital audio workstation is simple using this setup, and the data can be recorded into a MIDI track in real time. Ableton Live 11 Suite is excellent for this task as it has brilliant MIDI capability and the simple nature in which it can be used to integrate MIDI, Audio and live synthesis. Once the MIDI data is recorded into Ableton Live 11 Suite, there is a file to play, and can be listened to via a VST plug-in or hardware synthesiser. 

\subsection{Musicality of the data}\label{sec:music2}

A simple sine wave was used to hear the data initially, picked for its clarity and pure sound which makes it easy to hear definitive notes so focus can be placed on the relationship between each note. The raw data is atonal in nature, but differences between the quantum and classical random walks are clear, as is Grover's algorithm and how it changes over time and reduces down to just one note (finds the correct answer).

There are two main differences which can be heard in the music between the classical and the quantum walk. The first is that in the classical walk, the note only ever moves up or down by one step, making the melody sound quite similar to playing a scale. In the quantum walk, there are large jumps because the position of the walker is spread out over many positions at once. Another feature that can be heard is that in the quantum walk is that the notes move away from the starting point (the first note in the melody) faster than the classical walk. This is because of an important feature of the classical vs quantum walk - if you want to get somewhere fast, it helps to be quantum!

The first step into making the data more musical and `pleasant' to listen to was to quantise the notes to follow a musical key. C minor was chosen for this, specifically the harmonic minor scale. From a creative point of view, this felt musically fitting to the data from the laboratory. This choice was inspired by how the data sounded in its original atonal state - there was something about the movement of the notes that also inspired the choice to go with a minor key. After cycling through some different choices - C sounded the most appropriate. The next step is transposing the octave of the created melodies, in their raw form they are extremely low and are essentially inaudible. By shifting them up 3 octaves they are in a usable range.

\subsection{Composition Process}\label{sec:music3}

The quote below covers the composition process of the two finished pieces as documented from the composer Simon Small.   
\textit{
\begin{quote}
``After spending some time listening to the melodies created by the classical and quantum walks and Grover's algorithm, I found myself interpreting emotion and drawing inspiration from their raw forms. Grover's algorithm was a constant build to me, a crescendo, caused by the data honing in on its final destination. This led me to begin composing a track which in it's essence is one large crescendo, building intensity as the algorithm progresses. Listening to the random Walks together made me want to compose a track to show case their differences but also show how they can harmoniously play together. Keeping them as the main feature of the piece was important. The melody and harmony created really reminded me of early synthesiser music, specifically the works of Georgio Morodor. This influenced me to create a piece inspired by Morodo's synthesiser work and including drums from a drum machine opposed to an acoustic instrument.\\~\\
Aside from using the melody and harmony created by the experiment data, I also used fragments of the data with hardware synthesisers to create textural layers in the tracks. This was made incredibly simple by having the data available in the MIDI format. I wanted the entire piece to be influenced by data rather than a melody bolted on to an unrelated song.\\~\\
This was also achieved by using recorded audio samples from the physics Laboratory at the University of Bristol. Samples were recorded using an Apple iPhone by Euan Allen as I was unable to attend due to the Covid-19 pandemic. Equipment such as the lab's key card access point, cryostat, compressor and lab doors were recorded. They were used to create percussion, drones and textures within the compositions. This was an important creative point for me as again I wanted the whole composition to be created by and influenced by the data and the surroundings used in it's creation.''
\end{quote}
}

\newpage
\bibliographystyle{ieeetr}
\bibliography{bib.bib}

\end{document}